\documentclass[slac_one]{revtex4}
\usepackage{graphicx}
\usepackage{fancyhdr}
\begin{document}

\title{Development of a large area gas photomultiplier with GEM/$\mu$PIC} 

%

\author{H. Sekiya}
\affiliation{Kamioka Observatory, Institute for Cosmic Ray Research, University of Tokyo\\
456 Higashi-Mozumi, Kamioka, Hida, Gifu, 506-1205 JAPAN}

\begin{abstract}
We are developing a new photon detector with micro pattern gaseous detectors. 
A semitransparent CsI photocathode is combined with 
10cm$\times$10cm GEM/$\mu$PIC 
for the first prototype which is aimed for the large liquid Xe detectors. 
Using Ar+C$_2$H$_6$ (10$\%$) gas, we achieved the gas gain of $10^5$ which is 
enough to detect single photoelectron. We, then, irradiated UV photons from 
a newly developed solid scintillator, LaF$_3$(Nd), to the detector and 
successfully detected single photoelectron.
\end{abstract}

\maketitle

\thispagestyle{fancy}


\section{INTRODUCTION} 
In the last two decades, many gaseous photomultipliers with CsI photocathodes 
operated in both flushed gas mode and sealed gas mode 
have been developed and tested\cite{charpak,berskin,carlson}.
Moreover, recently, large area micro pattern gaseous detectors with avalanche 
multiplication structures, such as Micromegas\cite{megas}, GEM\cite{gem}, and
$\mu$PIC\cite{upic} have been developed and successfully operated. 
These devices with photocathodes can realize a low cost large area photon
detector with position sensitivity and can be applied to large 
astroparticle detectors.
In particular, the quantum efficiency of the CsI photocathode matches 
the liquid Xe scintillator, thus dark matter search via Xe is one of the 
first targets of this photon detector.

In this work, for the first step, we investigated the feasibility of
manufacturing a large size gaseous photon detector with CsI photocathode 
and evaluated its performance.

\section{THE DETECTOR}
\begin{figure}[htb]
\begin{center}
\includegraphics[width=10cm]{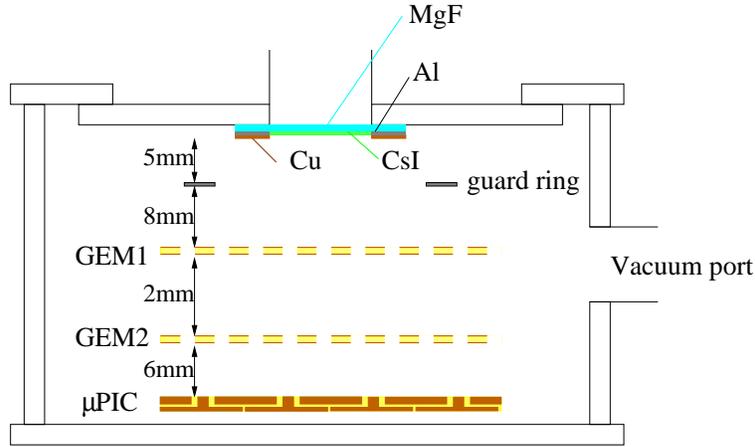}
\caption{Schematic drawing of the prototype detector. The size of GEMs
 and $\mu$PIC is 10cm$\times$10cm, while the diameter of the window is
 54mm and the diameter of the effective area of the CsI photocathode is 34mm.}
\label{fig:proto}
\end{center}
\end{figure}

 The configuration of the prototype detector is shown schematically in Fig. \ref{fig:proto}. 
2 GEMs and a $\mu$PIC were used for the charge amplification, which allows to suppress the
avalanche-induced photon and ion feedback and provide the high gain operation.

The GEM which was manufactured by SciEnergy Co., Ltd. is 10cm$\times$10cm size 
with $70\mu$m$\phi$ holes of 140$\mu$m pitch\cite{SMASH}, 
and the insulator is Liquid Crystal Polymer of 100$\mu$m thick.   
The $\mu$PIC is the standard type of 10cm$\times$10cm with 256 anode strips and
256 cathode strips developed at Kyoto University\cite{upic}.
Although $\mu$PIC has the very fine position sensitivity, 
for this prototype detector,
the cathode strips and anode strips were summed into 4$+$4 channels 
to reduce the number of the readout circuit.

 The MgF$_2$ window is 54mm$\phi$ and the thickness is 5mm.
The CsI photocathode was evaporated to the window by Hamamatsu Photonics
and the effective area is 34mm$\phi$.
In order to apply high voltage (-HV) to the photocathode
an Al electrode was also evaporated at the edge of the MgF$_2$ window.
The -HV is supplied to the electrode via contact Cu ring.  
Between the photocatode and the first GEM, a guard ring is placed
in order to make a uniform electric field  and to drift photoelectrons
to the first GEM.  

These devices are put in a stainless steal chamber with a
 gas/vacuum port, therefore, when the gas spoils, we can replace
the gas and we can also exchange the other components.
1 atm of Ar+10$\%$C$_2$H$_6$ mixture gas was loaded and the chamber
 was sealed during this measurement.

All the components given above are assembled in a nitrogen-sealed glove
box to avoid the deliquescence of the CsI photocathode. 

\section{UV PHOTON SCINTILLATOR}
The fluoride single crystals have a wide transparent range from the vacuum 
ultraviolet(VUV) to the infrared region, therefore, they are used in various
applications such as the window material, the host material for the 
short wavelength laser and so on. 
Recently, several studies have been made on fluoride crystals with dopant 
as scintillators that might find their applications in $\gamma$/x-ray 
detection\cite{yoshikawa}.  
Although light yields of these crystals are rather limited at this
point, because of their small light yields and easiness of handling, 
these crystals can be light sources for the prototype detector 
instead of liquid xenon.

It is known that the scintillation spectrum of the LaF$_3$(Nd) is 
agree with that of liquid Xe (the central wavelength is 172nm), however
the absolute light yield of  LaF$_3$(Nd) had not been measured yet.

\begin{figure}[htb]
\begin{center}
\includegraphics[width=8cm]{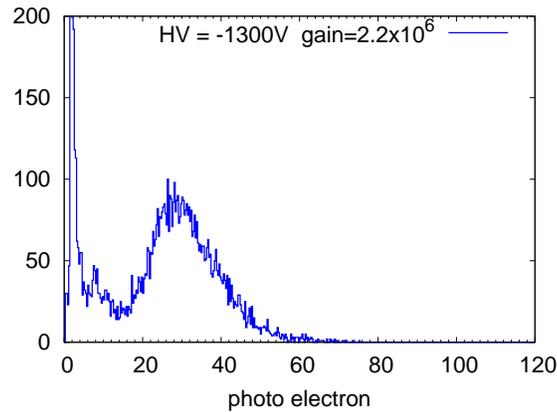}
\caption{Obtained spectrum of 5.5MeV $\alpha$ through the LaF$_3$(Nd). The R8778 was operated with the gain of 2.2$\times$10$^6$.}
\label{fig:LY}
\end{center}
\end{figure}

Accordingly, we measured the light yields of 15mm$\times$15mm$\times$15mm 
size LaF$_3$(Nd) for the first step. 
The $^{241}$Am (5.5MeV $\alpha$ source) was attached 
directly to one face of the crystal and the VUV sensitive PMT 
(Hamamatsu R8778) was attached to the opposite face. Other four sides
were covered with GORE-TEX$^{\mbox{\scriptsize{\textcircled{\tiny R}}}}$.
The obtained spectrum of 5.5MeV $\alpha$ through the LaF$_3$(Nd) is shown
in Fig. \ref{fig:LY}.

As the quantum efficiency of the R8878 at 172nm is 30$\%$, 
the photon yield of the LaF$_3$(Nd) was found to be 100 photons/5.5MeV $\alpha$.
We use this LaF$_3$(Nd)+ $^{241}$Am as a test light source for 
the prototype detector.

\section{RESPONSE TO THE VUV LIGHT SOURCE}  
\begin{figure}[htb]
\begin{center}
\includegraphics[width=6cm]{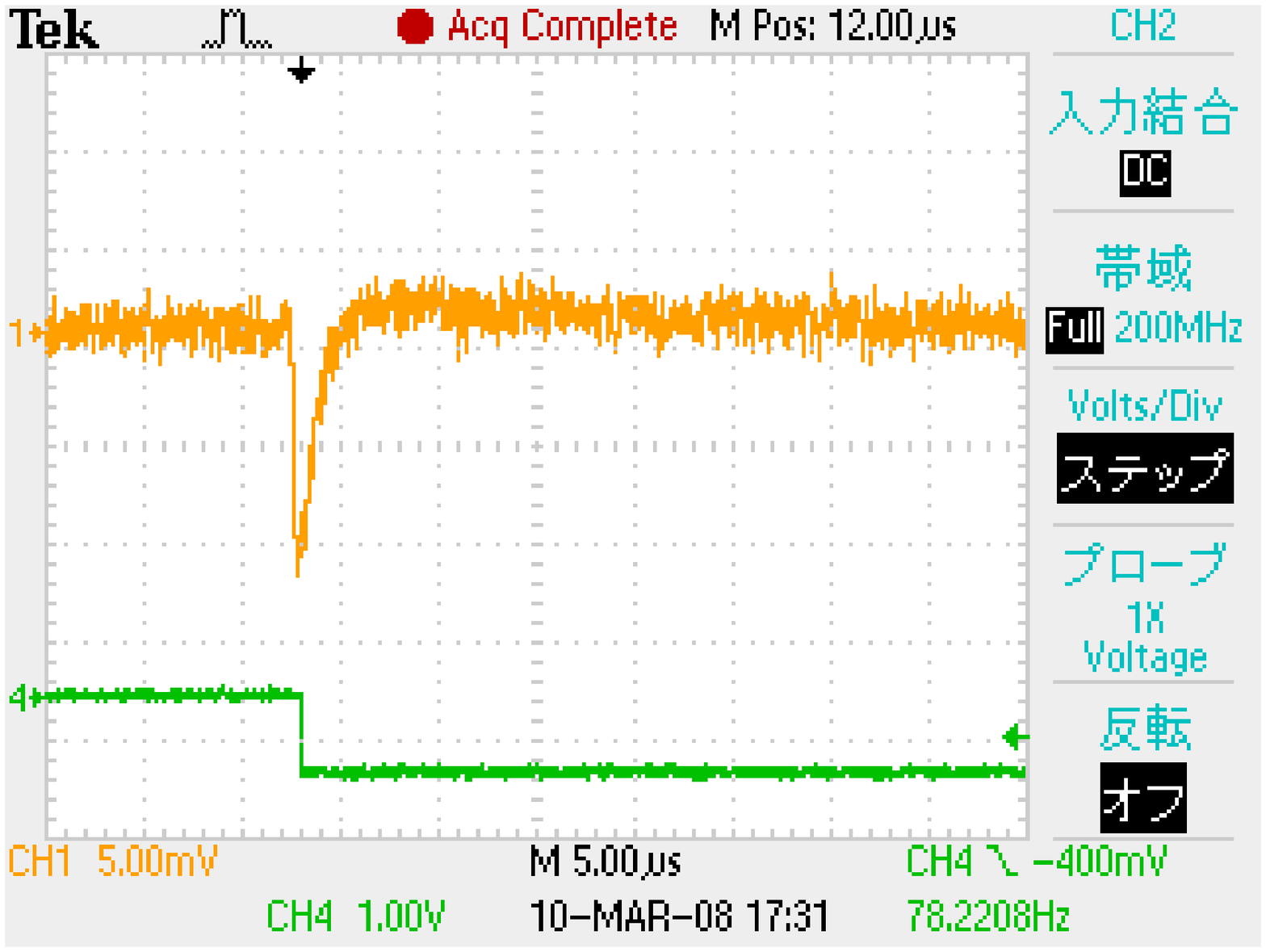}
\includegraphics[width=8cm]{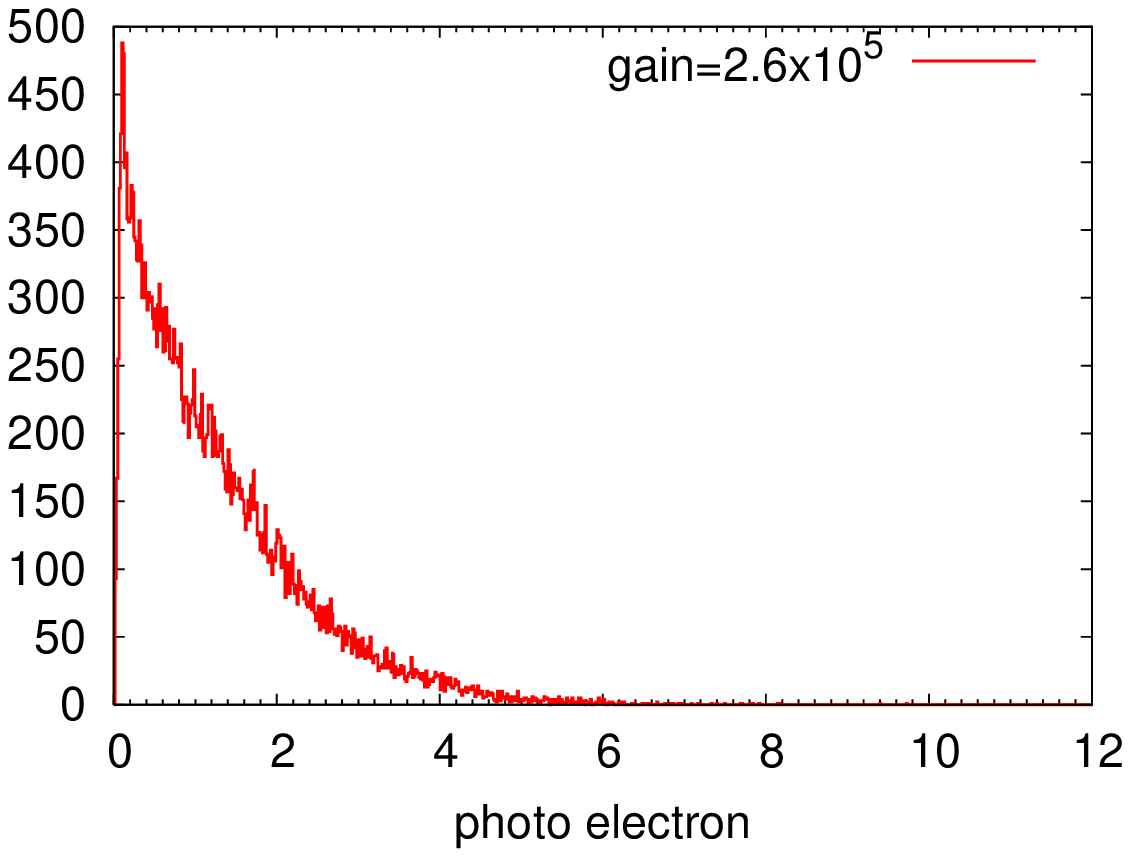}
\caption{Left: Typical output signal after the 0.1V/pC charge amplifier. Right: The obtained spectrum.  The detector was operated with the gain of 2.6$\times$10$^5$.}
\label{fig:wave}
\end{center}
\end{figure}

Next, we attached the light source to the prototype detector.
The detector was operated with the gas gain of 2.6$\times$ 10$^5$ stably
without any discharges more than one hour.
The typical output signal after the 0.1V/pC charge amplifier and the 
obtained spectrum are shown in Fig. \ref{fig:wave}.

It is confirmed that this detector has enough sensitivity to  
single photo-electron.
The quantum efficiency of this CsI photocathode is about 1$\%$
which is consistent with that in Hamamatsu's data sheet of the semitransparent 
CsI photocathode\cite{hpk}.

\section{CONCLUSION AND PROSPECTS}
A UV photon detector based a semitransparent CsI photocathode combined 
with large area GEMs and a $\mu$PIC has been reported.  
It is demonstrated that the detector has the ability of detecting single
photoelectron. The quantum efficiency is about 1$\%$ as expected.

In order to increase the quantum efficiency, a reflective type CsI 
photocathode (200nm-thick CsI evaporated to one side of the GEM) is
being developed and tested now.


\end{document}